%% file: main.tex
\documentclass[10pt,twocolumn]{article}
\pdfoutput=1
\usepackage{times}
\usepackage[linesnumbered,ruled,vlined]{algorithm2e}
\SetAlFnt{\footnotesize}
\SetAlgoSkip{}
\usepackage{multirow}
\usepackage{color,soul}
\usepackage{graphicx}
\usepackage{amsmath}
\usepackage{comment}
\usepackage{caption}
\usepackage{url}

\begin{document}


\title{Dynamic Scaling of Virtualized, Distributed Service Chains: \\A Case Study of IMS}

\author{Jingpu Duan$^1$, Chuan Wu$^1$, Franck Le$^2$, Alex Liu$^3$, Yanghua Peng$^1$ \\
\small {\em  $^1$The University of Hong Kong, Hong Kong\quad}
\small    {\em    $^2$IBM T.J. Watson Research Center, U.S.A.}\\
\small		{\em   $^3$Michigan State University, U.S.A.} 
}
\date{}
\maketitle

\begin{abstract}
	The emerging paradigm of network function virtualization advocates deploying virtualized network functions (VNF) on standard virtualization platforms for significant cost reduction and management flexibility. There have been system designs for managing dynamic deployment and scaling of VNF service chains within one cloud data center. Many real-world network services involve geo-distributed service chains, with prominent examples of mobile core networks and IMSs (IP Multimedia Subsystems). Virtualizing these service chains requires efficient coordination of VNF deployment across different geo-distributed data centers over time, calling for new management system design. This paper designs a dynamic scaling system for geo-distributed VNF service chains, using the case of an IMS. IMSs are widely used subsystems for delivering multimedia services among mobile users in a 3G/4G network, whose virtualization has been broadly advocated in the industry for reducing cost, improving network usage efficiency and enabling dynamic network topology reconfiguration for performance optimization. Our scaling system design caters to key control-plane and data-plane service chains in an IMS, combining proactive and reactive approaches for timely, cost-effective scaling of the service chains. We evaluate our system design using real-world experiments on both emulated platforms and geo-distributed clouds.

\end{abstract}

%
%



\input{introduction.tex}

\input{background.tex}
\input{motivation.tex}
\input{design-overview.tex}
\input{mechanism.tex}

\input{evaluation.tex}

\input{conclusion.tex}

\bibliographystyle{abbrv}

\end{document}

%% file: introduction.tex
\section{Introduction} \label{Introduction}

Traditional hardware-based network functions are notoriously hard and costly to deploy, maintain and upgrade. The recent trend towards Network Function Virtualization (NFV) promotes deploying software network functions in virualized environments over off-the-shelf servers, to significantly simplify deployment, maintenance and scaling at much lowered costs~\cite{nfv-website}. 

Despite the good news brought by NFV, many problems remain when applying it to practical network services, especially when scaling their service chains -- an ordered collection of virtual network functions (VNFs) that altogether compose a network service -- over a large geographical span. Existing work either focus on designing architectural improvement for NFV software, or creating management systems that scale VNF service chains on a single server cluster. These management systems are adequate for service chains that protect client-server based systems. For example, the entry service chain consisting of a firewall and an intrusion detection system (IDS) to access database and web services is typically deployed in the on-premise cluster/data center of the service provider. However, they cannot be directly applied to other service chains that provide inter-connection services for users residing widely apart, such as those in IP Multimedia Subsystems (IMS) and mobile core networks~\cite{3gpp-ims}~\cite{epc}. Putting such a service chain in a single datacenter would be unfavourable as compared to distributing its VNFs across several datacenters, to enable low-delay access of the users to the network functions. 
 
Designing a management system for scaling geo-distributed service chains imposes new challenges over the existing approaches. How to provision VNF instances on different datacenters, how to adjust the provisioning over time with variation of traffic, and what VNF instances a service chain should go through at different datacenters are all important problems that must be well addressed.

This paper presents \textit{ScalIMS}, a management system that enables dynamical deployment and scaling of VNF service chains across multiple datacenters, using the case of representative control-plane and data-plane service chains in an IMS. \textit{ScalIMS} is designed to provide good scaling performance (minimal VNF instance deployment and bounded flow end-to-end delays), using both runtime statistics of VNF instances and global traffic information. The IMS is chosen as the target platform to showcase \textit{ScalIMS} because of its important role in telecom core networks as well as the accessibility of open source software implementation of IMS network functions. \textit{ScalIMS} has two important characteristics that distinguish itself from the existing scaling systems: 

$\triangleright$ \textbf{Multi-datacenter scaling:} \textit{ScalIMS} deploys multiple instances of the same network function onto different data centers dynamically according to the traffic demand and distribution of users. In this way, the network paths taken by a service chain is optimized for small end-to-end delays and QoS guarantee of user traffic.

$\triangleright$ \textbf{Hybrid Scaling Strategy:} Most existing scaling systems~\cite{sherry2012making}~\cite{wood2007black}~\cite{gember2012stratos} rely on reactive scaling strategies, reacting to the change of runtime status of the network functions. \textit{ScalIMS} combines reactive scaling with proactive scaling, using predicted traffic volume based on history. This hybrid scaling strategy exploits all opportunities for timely scaling of network functions and improves the overall system performance. 

We evaluate \textit{ScalIMS} on both our own computing cluster and the IBM SoftLayer cloud. The experiment results show that \textit{ScalIMS} significantly improves QoS of user traffic compared with scaling systems that use only the reactive or proactive scaling approach. \textit{ScalIMS} achieves this improvement by either launching VNF instances timely or re-routing traffic to datacenters with redundant VNF instances, using almost $50\%$ less VNF instances. 
Even though we make some special design decisions for the IMS, the design principles of \textit{ScalIMS} can be easily applied to other NFV systems that benefit from service chain deployment in multiple datacenters.

%% file: background.tex
\section{Background} \label{Background}
\subsection{Overview of IMS}

\begin{figure}[!t]
        \centering
        \includegraphics[width=1\columnwidth]{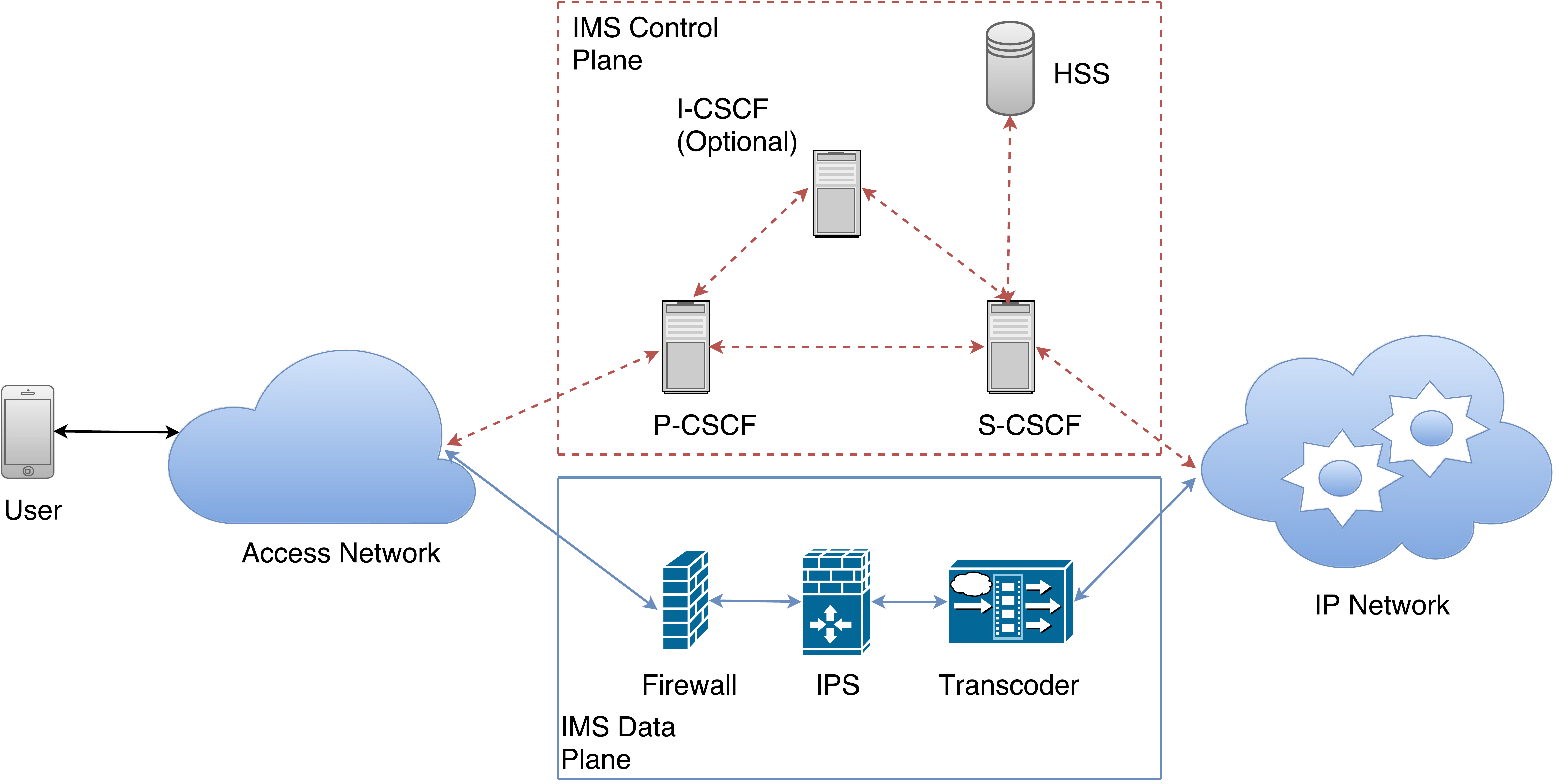}
        \caption{IMS: an architectural overview}
        \label{fig:IMS_architecture}
\end{figure}

An IP Multimedia System (IMS) is a core part in 3G/4G telecom networks (e.g., 3GPP, 3GPP2)~\cite{umts}~\cite{lte}, responsible for delivering multimedia services (e.g., voice, video, messaging) over IP networks. IMS is a fully standardized solution for multimedia service delivery across the telecom industry~\cite{3gpp-ims}, as compared to its proprietary alternatives (e.g., Skype, Facetime). It is a complex system consisting of multiple service chains. We investigate two most important service chains as follows.
An illustration is given in Fig.~\ref{fig:IMS_architecture}.

\noindent\textbf{Control Plane Service Chain} includes three main network functions, {Proxy-CSCF (P-CSCF)}, {Interrogating-CSCF (I-CSCF, which is optional)}, and {Serving-CSCF (S-CSCF)}, 
 which user call requests traverse for user registration and authentication, call setup and charging. The network functions rely on the IETF Session Initiation Protocol (SIP)~\cite{sip} to interoperate with external terminals (i.e., mobile users) in the Internet. The three types of network functions act as different roles of SIP proxies and servers. In addition, S-CSCF relies on an external server called Home Subscriber Server (HSS), which is a user database that contains the subscription-related information for S-CSCF to query.

\noindent\textbf{Data Plane Service Chain} contains a sequence of network functions that the actual multimedia traffic between users traverses, for security (e.g., firewall, deep packet inspection, intrusion detection), connectivity (e.g., NAT, IPv4-to-IPv6 conversion), quality of service (e.g., traffic shaping, rate limiting, ToS/DSCP bit setting), and media processing (e.g., transcoding). While 3GPP has standardized the IMS control plane for interoperability reasons, the exact set of deployed network functions for the data plane varies per operator.

The two service chains collectively handle two important procedures of an IMS system, which are user registration and session origination. 
When a user would like to make a call, he first announces his IP address to the IMS by initiating a SIP REGISTRATION transaction over the control plane. When the registration is done, S-CSCF temporarily stores this (user, IP address) binding for session origination. When a registered caller initiates a SIP INVITE transaction to a callee over the control plane, i.e., session origination starts, S-CSCF uses the binding saved during user registration procedure to retrieve the IP address of the callee and routes the SIP message to the callee. After the callee responds, a call is successfully set up. Subsequent media flows between the caller and the callee are routed through data plane service chain. When the call is finished, a SIP BYE transaction between the caller and the callee is carried out to close the call over the control plane.

\subsection{Related Work}


Performance and scalability of using VNFs in the place of hardware middleboxes have been the focuses of investigation in the existing literature on NFV.

\noindent \textbf{Performance.} Running VNF software (e.g., data plane packet processing software~\cite{kohler2000click}) on VMs incurs significant context switching cost~\cite{rizzo2013speeding}, limiting the maximum throughput of a VNF. To solve this problem, systems such as ClickOS~\cite{martins2014clickos} and NetVM~\cite{hwang2015netvm} map packets directly from NIC receive queues to the memory and fetch packets directly from user space~\cite{dpdk}~\cite{rizzo2012netmap}, which greatly improve the throughput of VNF. 

\noindent \textbf{Scalability.} The existing studies have investigated scaling of VNFs in a single server, in a computing cluster and in a datacenter. RouterBricks~\cite{dobrescu2009routebricks}, xOMB~\cite{anderson2012xomb} and CoMB~\cite{sekar2012design} focus on scaling VNFs in a single server for better performance. 
Their designs typically rely on customized VNF software design and may lack generality and flexibility. 
 E2~\cite{palkar2015e2} scales VNF instances within a computing cluster connected through SDN enabled switches. 
  Both Stratos~\cite{gember2012stratos} and Slick~\cite{anwer2015programming} study how to scale VNF in a SDN-enabled datacenter. Stratos~\cite{gember2012stratos} scales VNF instances by jointly considering VNF placement and flow distribution within a datacenter, using on-demand VNF provisioning and VM migration to mitigate hotspots. Slick~\cite{anwer2015programming} focuses on providing a programing interface for managing VNF instances in a datacenter. To our knowledge, there do not exist management frameworks that scale VNF service chains across geo-distributed datacenters. 
And existing works can't be directly extended to multi-datacenter scaling. The primary reason is that SDN~\cite{mckeown2008openflow} is extensively used by existing approach to control routing, scaling and loadbalancing within a single datacenter, but it is extremely hard for a SDN controller to set up flows on other datacenters, which incurs too much delay and hurts flow performance ~\cite{tootoonchian2012controller}. One possible way is to deploy one scaling system on each datacenter, but there's no existing works to coordinate the behavior of each independent scaling system. However, scaling systems on different datacenters need to agree on complicated task such as coordinated VNF instance provision and collective flow routing. And all these problems call for an efficient design of a multi-datacenter scaling system.

  \textit{ScalIMS} uses similar methodologies as in ~\cite{gember2012stratos}~\cite{anwer2015programming}~\cite{wood2007black} when scaling NFV service chains within a datacenter, but adopts a novel distributed flow routing framework and proactive scaling to scale NFV service chains across multiple datacenters.

%% file: motivation.tex
\section{Motivation and Design Highlights} \label{Motivation}

\noindent \textbf{Motivation.}
The key motivation of this study is the fact that there are no NFV scaling systems designed to operate under a multi-datacenter environment, given the following new challenges compared with scaling service chains within a single datacenter.

\begin{figure}[!tbp]
      \includegraphics[width=\columnwidth]{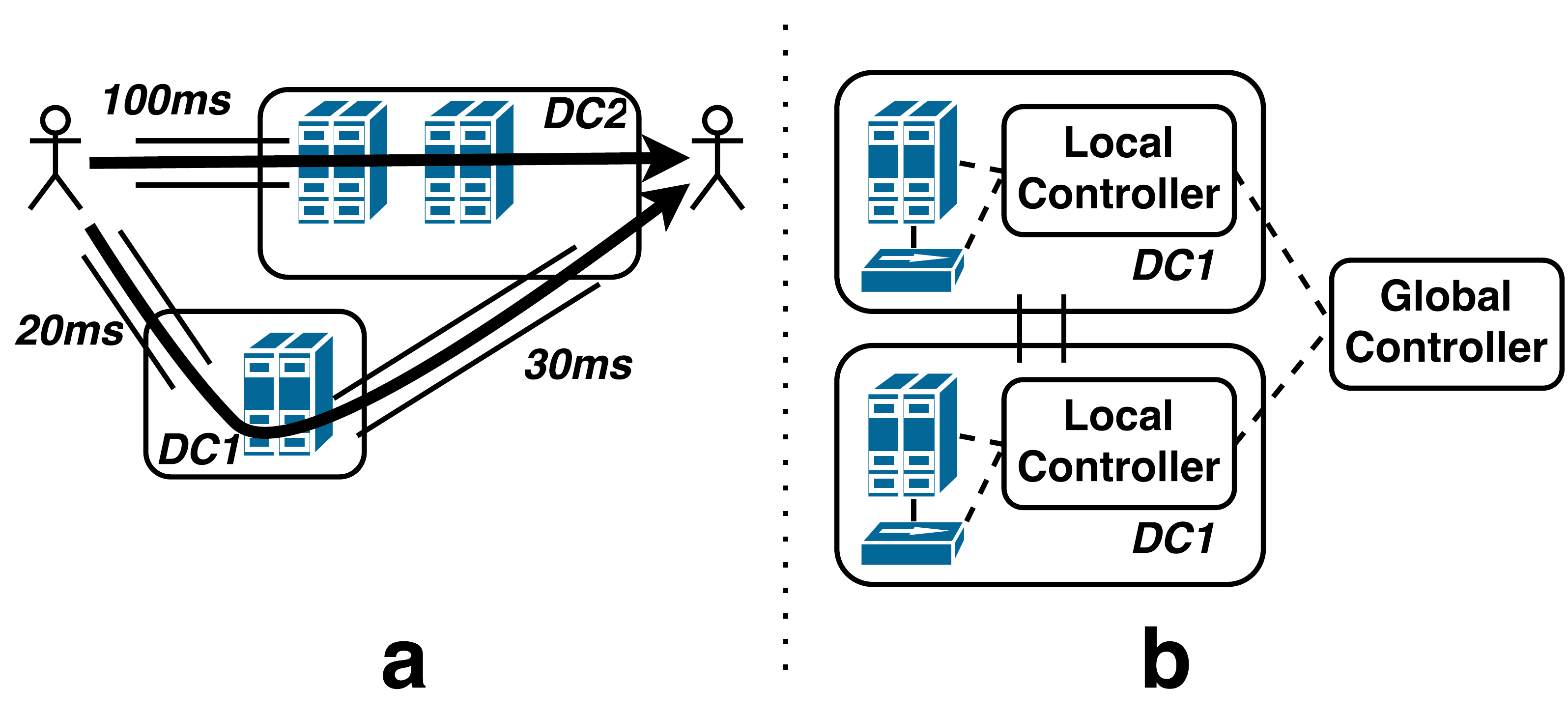}
    \caption{(a) Service chain path in IMS system: an example; (b) ScalIMS: an overview}
    \label{fig:system-overview}
\end{figure}

The {\em first} challenge is how to decide the service chain paths, {\em i.e.}, the datacenters where instances of VNFs in a service chain are to be deployed. The deployment has an important impact on service quality of user traffic along the chain. In Fig.~\ref{fig:system-overview}(a), traffic of IMS system user can go through either of the two service chain paths (containing two VNFs each) to reach another IMS user. Choosing the lower path incurs a 50ms end-to-end delay and choosing the upper path incurs a 100ms delay. A multi-datacenter NFV scaling system should constantly update the service chain paths so that a good service quality can be guaranteed for user traffic all the time.

The {\em second} challenge lies in that decisions on scaling of network functions ({\em i.e.}, adding/removing instances of one VNF upon traffic increase/decrease) and deployment of service chain paths are coupled in the multi-datacenter scenario. If a service chain path is suffering from overloading, instead of launching new network function instances directly on the same datacenters, the scaling system may search for available network function instances on other datacenters and move the service chain path to those datacenters. 

The {\em third} challenge is the quest for distributed traffic routing. Using SDN to control flow routing in a data center is a common approach adopted by the existing scaling systems. However, if a service chain path goes through multiple datacenters, it becomes impossible for a single SDN controller to control the end-to-end route. Multiple SDN controllers on a service chain path should work in coordination, upon constant updates of the path, to correctly route user traffic towards the destination.

\vspace{1mm}
\noindent\textbf{Highlights of {\em ScalIMS}.}
The overall architecture of \textit{ScalIMS} is shown in Fig.~\ref{fig:system-overview}(b), and we have made the following design decisions.

{\em First}, we adopt a hybrid scaling strategy combining proactive scaling and reactive scaling for both control plane and data plane service chains. We divide the system time into \textbf{scaling intervals}. At the end of each scaling interval, proactive scaling is carried out taking as input global information including predicted workload, inter-datacenter delays and the current VNF deployment (the numbers of instances of each VNF on each datacenter),  generating decisions on VNF instance scaling and service chain path deployment simultaneously for the next scaling interval. Reactive scaling produces scaling decisoins of each VNF based on runtime statistics of each instance locally. It compensates for the inaccuracy of workload prediction in proactive scaling, improving system performance under unpredictable traffic volume surges.

{\em Second}, \textit{ScalIMS} enables a synergy of global and local controllers (Fig.~\ref{fig:system-overview}b), to best execute the hybrid scaling strategy. For proactive scaling, the global controller coordinates with all local controllers: it collects input data from each local controller, executes the proactive scaling algorithm, generates scaling/deployment decisions, and broadcasts the decisions to local controllers. Each local controller executes the received decisions by launching new VNF instances and adjusting service chain paths. For reactive scaling, a local controller collects runtime statistics from each VNF instance running on the same datacenter and produces reactive scaling decision. For flow routing, a local controller uses flow tags and service chain paths received from the global controller to determine the VNF instances that a flow should traverse.

{\em Third}, we bound the end-to-end delay on service chain paths to guarantee a good end-to-end performance of flows. We also make some design decisions specially tailored for an IMS system, {\em e.g.}, SIP message content manipulation and network address translation. Similar design philosophies can nevertheless be applied in other service chain systems as well.

%% file: design-overview.tex

\section{Scaling of CP Service Chain}
\label{System-Design}

In this section, we present the detailed design of \textit{ScalIMS} in deployment and scaling of the control plane (CP) service chain.



\subsection{Preliminary}

\textbf{Entry Datacenter Binding:} Each user of the system is bound to an {\em entry datacenter}. If user $a$ sends a flow to user $b$, then the flow will enter the service chain from user $a$'s entry datacenter and exit from user $b$'s entry datacenter, before being sent to user $b$. User $b$'s entry datacenter is also referred to as user $a$'s {\em exit datacenter}. We use a simple location service to assign a user to his entry datacenter: we map a user's IP address to a geographical location by querying an IP-location database ({\em e.g.}, IP location finder~\cite{iplocation}), and then assign a datacenter that is closest to user's current location as his entry datacenter.

\noindent \textbf{Fixed CP Service Chain Placement:} The CP service chain of an IMS consists of P-CSCF and S-CSCF. We use a fixed placement strategy by deploying P-CSCF instances on every entry datacenter and S-CSCF instances on a fixed datacenter. The delay between the datacenter where we place S-CSCF instances and other datacenters should fall within an acceptable range. 

The reason why we adopt such a fixed placement strategy is that S-CSCF instances relies on external appliances to work. First, S-CSCF constantly queries the HSS database for user information. In addition, S-CSCF instances are usually built as a stateless network function by storing user location information (a temporary mapping between user name and user registration IP address) in a memcached cluster. The stateless design improves load balancing among S-CSCF instances, but also implies that even if we spread S-CSCF instances across different datacenters, each S-CSCF instance still needs to access a central HSS server and a memcached cluster to process most of the SIP transactions. This fact urges us to place S-CSCF instances together with the HSS server and memcached cluster in the same datacenter. P-CSCF instances act as relay points to access S-CSCF instances. By placing them on each entry datacenter, a user's flow can always access a P-CSCF instance on the closest datacenter.

We use a mixed proactive and reactive strategy for deciding the number of instances of P-CSCF and S-CSCF to deploy in the respective data centers.

\subsection{Proactive Scaling}
\label{sec:CP_proactivescale}

\begin{figure}[h]
        \centering
        \includegraphics[width=1\columnwidth]{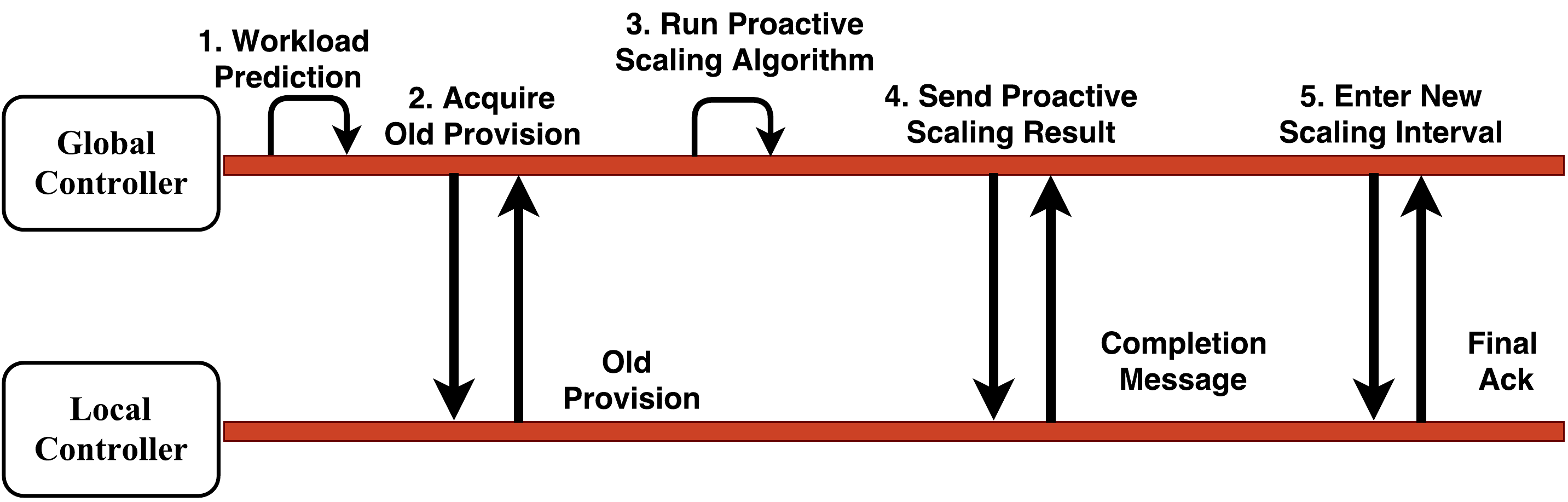}
        \caption{Proactive scaling workflow.}
        \label{fig:proactive-scaling}
\end{figure}

Proactive scaling is executed in each scaling interval according to the workflow in Fig.~\ref{fig:proactive-scaling}. The global controller and each local controller work in a request-response manner. In view of possible message loss on the way, the global controller sets a re-transmision timer that re-transmits a request after 500ms, if no response is received.



\noindent\textbf{1. Workload Prediction:}
Workload along a CP service chain is described by the number of SIP transactions carried out on each entry-exit datacenter pair every second. When a SIP transaction finishes, the S-CSCF instance involved uses the location service to determine which entry-exit pair this transaction belongs to (according to the IP addressed of the caller and the callee). Each S-CSCF instance keeps a record of the number of SIP transactions on each entry-exit datacenter pair and reports this number to the local controller in its datacenter every second . The local controller accumulates CP workload for 5 seconds before relaying it to global controller.

At the end of the current scaling interval $t$, the global controller predicts the workload $\hat{u}_{t+1}$ in the next scaling interval using historic data in several past intervals (10 as in our experiments) based on auto regression~\cite{wood2007black}~\cite{box2015time}:

\begin{equation}
\hat{u}_{t+1} = \mu + \phi(u_t-\mu)
\end{equation}



\noindent Here $\mu$ is the mean of the historic values for several past scaling intervals, $u_t$ is the average value of the collected CP workload in the current scaling interval and $\phi$ is a parameter (it captures the variation of the data series formed by received workload values~\cite{wood2007black}~\cite{box2015time}). The workload between each entry-exit datacenter pair is being predicted this way.

\noindent\textbf{2. Aquiring Current Provision:} The global controller then broadcasts a message to local controllers, asking them to send information of the current VNF instance provisioning ({\em i.e.}, the number of instances of each VNF provisioned in the respective datacenter). Upon receiving this message, a local controller stops its reactive scaling process (to be discussed next) so that it does not interfere with proactive scaling, and then responds to the global controller by sending its current provisioning information.

\noindent\textbf{3. Compute Proactive Scaling Results:} After receiving information of current provisioning from all local controllers, the global controller computes the number of P-CSCF and S-CSCF instances to be deployed in the respective datacenters in the next scaling interval. Since all S-CSCF instances are placed in the same datacenter, the number can be decided by dividing the total predicted traffic volume between all pairs of entry-exit datacenters by the processing capacity of one instance. The number of P-CSCF instances to be deployed in a datacenter can be computed by dividing the overall predicted volume of traffic flows, that use the datacenter as either the entry point or the exit point, by the processing capacity of one instance.


\noindent\textbf{4. Broadcast Proactive Scaling Result:} The global controller then broadcasts the computed new deployment numbers to local controllers. A local controller sends a completion message to the global controller, after applying the new provisioning result by creating new VNF instances (scale-out) or enqueueing unused VNF instances to tail of buffer queue (scale-in).


\noindent\textbf{5. Enter New Scaling Interval: } After the global controller receives completion messages from all local controllers, it broadcasts an ``enter new scaling interval message'' to all local controllers. After getting this message, a local controller increments its scaling interval index by 1, and shuts down some VNF instances from the head of buffer queue. Then the local controller sends a final acknowledgement to the global controller. After receiving all final acknowledgements from all local controllers, the global controller increases its scaling interval by 1.


In each datacenter, a double-ended buffer queue is maintained to temporarily hold unused VNF instances, with one queue for one type of VNF. When a VNF instance is to be removed, instead of directly shutting it down, local controller tags it with the current scaling interval index and enqueues it to the tail of the buffer queue. Once a VNF instance is enqueued into the buffer queue, no more flows will be routed to it. Whenever more VNF instances are to be established, if there are available instances in the buffer queue, buffered instances from the tail of the respective queue will be poped out and transformed back to working instances as much as possible. The purpose is to avoid frequent creation/deletion of VNF instances under workload fluctuation. Unused buffered VNF instances are destroyed after $\tau$ scaling intervals ($\tau$ is set to 10 scaling intervals in our implementation): when the local controller in the datacenter receives an ``enter new scaling interval message'', it checks from the head of each buffer queue whether an instance has been enqueued $\tau$ intervals ago; if so, the VNF instance at the head will be dequeued. Dequeuing repeats until all unused instances for $\tau$ intervals are cleaned. These dequeued VNF instances gets destroyed by local controller when there's no active traffic on them.

\subsection{Reactive Scaling}
\label{sec:CP_reactivescale}


A statistics reporting agent is running on each VNF instance and reports to the local controller runtime statistics for CPU usage, memory usage and input packet number every second. The local controller maintains time series for each type of metrics of each VNF instance. It decides whether CPU/memory/network is overloaded during the past several seconds (5 seconds as in our experiments) by comparing the respective runtime statistics with a threshold. 
If overload is persistently identified for at least two types of metrics ({\em e.g.}, CPU usage and network, in order to eliminate false alarms) for several consecutive seconds (5 seconds as in our experiments), then that VNF instance is reported as overloaded. Its state will be transferred from ``normal" to ``overload". The local controller will avoid routing new traffic flows ({\em i.e.}, new calls) to overloaded instances if there are available ``normal" instances. In a datacenter, if the states of a majority of VNF instances with the same type are ``overload'', scale-out is triggered by adding one new instance of that type. Note that no scale-in decisions (i.e., removing instances) are made reactively, which is solely handled by proactive scaling.

\subsection{Message Routing on Control Plane}
\label{sec:message-routing-on-control-plane}


When a user connects to an IMS, he first issues a query to a global DNS server, which determines the entry datacenter of this user using the location service and obtains the IP address of an available P-CSCF instance (non-overloaded) by querying a local DNS server located in the entry datacenter. A P-CSCF instance learns the IP addresses of several available S-CSCF instances (non overloaded) through DNS query as well. The DNS record of each CP VNF instance is updated by the local controller in the datacenter hosting the VNF instance.
P-CSCF instance maintains several connections to up-stream S-CSCF instances and evenly balance up-stream requests among these connections. It regularly (every 30s in \textit{ScalIMS}) updates these connections by issuing new DNS query to obtain an updated view of S-CSCF instances.

The establishment of a call involves that the caller sends out a SIP INVITE message and the callee responds with a SIP OK message. The content of these two messages are modified by a P-CSCF instance in order for the caller/callee to learn a destination IP address on his entry datacenter. A SIP INVITE message carries the caller's source IP address, receive port and send port. When a SIP INVITE message passes through callee's entry datacenter, the source IP field is modified to an IP address located on the callee's entry datacenter, so that the callee learns about a destination IP address located on callee's entry datacenter. A similar process is applied to SIP OK message as well. Besides SIP message modification, a P-CSCF instance sends two mappings (shown in Table~\ref{mappings}) to its local controller after it has received the SIP OK message. The local controller saves these two mappings for use when it processes the data plane traffic flows. Details will be discusses in Sec.~\ref{sec-dp-scaling}.

\begin{table}[t]
\centering
\resizebox{\columnwidth}{!}{%
\begin{tabular}{| l |p{0.75\columnwidth}|}
\hline
\multirow{2}{*}{\begin{tabular}[c]{@{}l@{}}Caller Entry DC\\ Controller\end{tabular}} & 1. (caller IP, caller send port)$\rightarrow$(callee IP)                   \\ \cline{2-2}
                                                                                & 2. (callee IP, callee send port)$\rightarrow$(caller entry IP, caller IP, caller receive port) \\ \hline
\multirow{2}{*}{\begin{tabular}[c]{@{}l@{}}Callee Entry DC\\ Controller\end{tabular}} & 3. (callee IP, callee send port)$\rightarrow$(caller IP)                   \\ \cline{2-2}
                                                                                & 4. (caller IP, caller send port)$\rightarrow$(callee entry IP, callee IP, callee receive port) \\ \hline
\end{tabular}
}
\caption{Mappings Saved On Local Controller of the Entry Datacenter}
\label{mappings}
\end{table}

%% file: mechanism.tex

\section{Scaling of DP Service Chain}

\label{sec-dp-scaling}
We next present the detailed design of {\em ScalIMS} in deployment and scaling of the data plane (DP) service chain. DP shares with CP the same reactive scaling mechanism discussed in Sec.~\ref{sec:CP_reactivescale}. For proactive scaling, DP follows the same  proactive scaling workflow in Sec.~\ref{sec:CP_proactivescale}, with some minor differences. 


{\em First}, DP proactive scaling scheme, running in Step 3 of Fig.~\ref{fig:proactive-scaling}, not only decides how VNF instances are provisioned in each datacenter, but also updates the service chain path between each entry-exit datacenter pair, by running the DP proactive scaling algorithm (see Sec.~\ref{sec:dp-proactive-scaling-alg}). We refer to VNFs in a DP service chain as {\em stages} of the chain and index them following the order of the VNFs in the chain. For example, DP service chain used in \textit{ScalIMS} contains firewall (stage 1), IDS (stage 2) and transcoder (stage 3). Since \textit{ScalIMS} manages multiple datacenters, we define a {\em service chain path} between each pair of entry-exit datacenters to be a path of datacenters. For a service chain with $m$ stages, a service chain path contains $m+2$ datacenters. The $0$th and $m+1$th datacenter are entry datacenter and exit datacenter of the entry-exit datacenter pair associated with the service chain path. The $i$th datacenter ($i=1, ..., m$) hosts instances of stage $i$ VNF in the service chain. Stage $1$ and stage $m$ could be hosted on other datacenters other than entry and exit datacenter. For example, a service chain path of $(0, 0, 1, 1, 2)$ shows that datacenter 0 is the entry datacenter, stage 1 to stage 3 of the service chain are hosted on datacenters 0, 1, and 1, respectively, and the exit datacenter is datacenter 2. Note that even if instances of a VNF may be deployed on different datacenters, we maintain only one service chain path for each service chain between each entry-exit datacenter pair in each single scaling interval, for path computation efficiency.

{\em Second}, Besides provisioning VNF resources, local controller also acts as an SDN controller that manages DP flow routing within local controller's datacenter. When the local controller receives DP proactive scaling results in step 4 of Fig.~\ref{fig:proactive-scaling}, the local controller applies the DP VNF provision decisions (i.e., add/remove instances) immediately, but saves the new DP service chain paths and uses them for routing only after receiving an ``enter new scaling interval'' message in step 5 of Fig.~\ref{fig:proactive-scaling}. 
This 2-stage update is used to guarantee consistency for distributed flow routing (see Sec.~\ref{Inconsistency}). 

{\em Third}, the proactive scaling algorithm takes as input predicted workload, predicted inter-datacenter network delay, and the current DP VNF deployment. Workload along a DP service chain is described as the number of packets transmitted over each entry-exit datacenter pair every second. The local controller acquires input DP workload using workload measuring OpenFlow rules installed on the entry switch. The local controller reports DP workload measurements to the global controller every second. Inter-datacenter delays are constantly measured by local controllers through a ping test among each other. Each local controller reports the ping delay with the other local controllers to the global controller every second. At the end of the current scaling interval, the global controller perdicts workload and delays in the next interval using the same approach as discussed in Sec.~\ref{sec:CP_proactivescale}.


\subsection{DP Proactive Scaling Algorithm}
\label{sec:dp-proactive-scaling-alg}

The DP proactive scaling algorithm is presented in Alg.~\ref{algo:dpalg}. It greedily serves the predicted workload of each entry-exit datacenter pair $p$ using the current VNF instance provisioning, as long as the capacity is still sufficient and the end-to-end delay threshold is still guaranteed (lines 2-4). Otherwise, a new service chain path for the entry-exit datacenter pair is computed using Alg.~\ref{algo:pc}. If there is not enough VNF capacity to serve the predicted workload on $p$'s new service chain, $\lceil (Q-Q')/C \rceil$ new VNF instances are created for the respective VNF in the respective datacenter (whose capacity is in shortage), where $Q$ is $p$'s predicted workload, $Q'$ is the total capacity of the VNF in the respective datacenter and $C$ is the processing capacity of each instance of that VNF (lines 5-9). For an entry-exit datacenter pair $p$ with the same entry and exit datacenter, the entire service chain path of $p$ is always deployed in this datacenter. The algorithm processes these entry-exit datacenter pairs at last so that the VNF capacity can be utilized more efficiently (lines 10-14).
 When workload of all entry-exit pairs is packed, scale-in is carried out to remove 
 un-used VNF instances in the corresponding datacenter. 

\begin{algorithm}[!t]
\KwIn{Predicted delay between each datacenter pair, predicted workload of each entry-exit datacenter pair, current VNF instance provision, current service chain paths}
\KwOut{New VNF instance provision, new service chain paths}
{Compute total available processing capacity of instances of each VNF in each datacenter}\;
\ForEach{entry-exit datacenter pair $p$, $p.entry \neq p.exit$} {
 	\If{there is enough VNF capacity on $p$'s current service chain path and the end-to-end delay threshold is not violated along $p$'s current path}{
		{use $p$'s current service chain path and decrease predicted workload from available processing capacities of VNFs on $p$'s new path}\;
	}
}
\ForEach{$p$, $p.entry \neq p.exit$ $\&\&$ $p$'s new service chain path has not been determined} {
 	{compute a new path for $p$ using Alg.~\ref{algo:pc}}\;
	\If{there is not enough capacity on $p$'s new path}{
		{scale out by creating new VNF instances sufficient to serve $p$'s workload}\;
	}
	{decrease predicted workload from available processing capacities of VNFs on $p$'s new path}\;
}
\ForEach{$p$, $p.entry = p.exit$} {
 	{use $p$'s current service chain path}\;
	\If{there is not enough capacity on $p$'s new path}{
		{scale out by creating new VNF instances sufficient to serve $p$'s workload}\;
	}
	{decrease predicted workload from available processing capacities of VNFs on $p$'s new path}\;}
{scale in by enqueuing un-used VNF instances to the respective buffer queues}\;
\caption{DP Proactive Scaling Algorithm}
\label{algo:dpalg}
\end{algorithm}

\subsection{Path Computation Algorithm}
\label{sec:pathcomp}

Besides conforming to the end-to-end delay requirement, a DP service chain path should meet two restrictions: (1) each datacenter in a service chain path should host at least one VNF in the respective service chain; 
(2) a service chain path should be loopless i.e., for two VNF stages $x<y$, if instances of both stage $x$ and stage $y$ are placed on datacenter $i$, then instances of stage $z$, $x<z<y$, must also be placed in datacenter $i$, as otherwise a routing loop is created. 

\begin{algorithm}[!t]
\KwIn{Predicted inter-datacenter delays, $p=(entry, exit)$, $p$'s predicted workload, $p$'s current service chain path, current available processing capacities of VNF instances, the service chain of $m$ stages, $n$ datacenters}
\KwOut{$p$'s new service chain path}
{$minProvPath = p$'s current path}\;
\For{$v = 0, ..., exit-1, exit+1, ..., n-1$}{
	{$record[0] = entry$, $record[1] = v$, $record[m+1] = exit$}\;
	\For{$x = 2, ..., m$}{
		{find out datacenter $v_1$ ($v_1 \neq exit$) that has the largest available capacity for stage-$x$ VNF}\;
		{$record[x] = v_1$}\;
  		\If{there is path loop on $record$}{
   			{eliminate loop by adjusting $record$}\;
  		}
		{$path[0, ..., x] = record[0, ..., x]$}\;
		{$path[x+1, ..., m+1] = exit$}\;
		\If{$path$ leads to a smaller number of new VNF instances to create for hanlding predicted workload than minProvPath and satisfies end-to-end delay requirement}{
		 minProvPath = path
		}
 	}
	{$path[0]=entry, path[1]=v, path[2, ..., m+1] = exit$}\;
	{check whether $path$ should be assigned to $minProvPath$ as in lines 11-12}\;
}
{$path1[0]=entry, path1[1, ..., m+1] = exit$}\;
{$path2[0, ..., m]=entry, path2[m+1] = exit$}\;
{check whether $path1$ or $path2$ should be assigned to $minProvPath$ as in lines 11-12}\;
\If{$minProvPath$ violates end-to-end delay requirement}{
 {find out the shortest-delay datacenter path between entry and exit}\;
 {run stage placement algorithm and assign the found service chain path to $minProvPath$}\;
}
\Return{$minProvPath$}\;
\caption{Path Computation}
\label{algo:pc}
\end{algorithm}

Alg.~\ref{algo:pc} presents the algorithm to compute a good service chain path for a given service chain between a given entry-exit datacenter pair, that aims to minimize the number of new VNF instances to be created for handling the predicted workflow between the entry-exit datacenter pair, while satisfying the end-to-end delay requirement (lines 10 $\sim$ 11). If the DP service chain contains $m$ stages and there are $n$ datacenters, exhaustive search to identify such a service chain path requires $O(m^n)$ running time. Instead, Alg.~\ref{algo:pc} seeks to optimistically find a good path in $O(mn)$ time.

In Alg.~\ref{algo:pc}, array $record$ retains the path under investigation (line 3). The search starts by looping through all datacenters except the exit to decide one for hosting VNFs of stage $1$ 
(lines 2 $\sim$ 3). For each subsequent stage of the service chain, a datacenter (except the exit) with the largest available capacity of VNF instances of that stage is selected (lines 4 $\sim$ 6). We might have created a loop in the datacenter path. If so, we eliminate the loop by the method to be discussed in \textbf{Loop Elimination} below (lines 7 $\sim$ 8). Once the datacenter to host the stage-$x$ VNF is determined, a candidate path is produced (lines 9 $\sim$ 13): we compute the number of new VNF instances to be created along the path, beyond the existing available capacities, for serving the predicted workload between the entry-exit pair,  and end-to-end delay of the path; if the path incurs less addition of new VNF instances and satisfies the delay requirement than the current best candidate path, we retain it in $minProvPath$ (lines 12-13). The algorithm also checks some candidate paths that are not generated by the search loop (lines 13 $\sim$ 17). Finally, it is possible that all candidate paths found so far fail to satisfy the delay requirement. If so, we compute the datacenter path with the shortest end-to-end delay ({\em e.g.}, using a shortest path algorithm) and run the {\bf Stage Placement Algorithm} below to produce the service chain path (lines 18-20).

\textbf{Loop Elimination:} 
 A routing loop may be introduced in lines 5 $\sim$ 6. For example, suppose the {\em record} contains $(1, 2, 4, ...)$, meaning entry datacenter is datacenter 1, instances of stage-1 VNF are placed on datacenter 2, and instances of stage-2 VNF are placed on datacenter 4. If datacenter $2$ is selected to host the stage-$3$ VNF, then the {\em record} becomes $(1, 2, 4, 2, ...)$ and a loop is created. We compare two approaches for eliminating the loop. The first is that we deploy VNF instances of stages involved in the loop in the same datacenter at the start and the end of the loop, {\em e.g.}, deploy stage $2$ VNF on datacenter 2 instead of datacenter 4 to make the path $(1, 2, 2, 2, ...)$. The second option is that we always choose the datacenter with the largest available capacity for hosting a stage, without adding a routing loop to the existing path. Suppose datacenter $3$ has the second largest capacity of stage-$3$ VNF. Then we choose datacenter 3 instead of datacenter 2 to host stage-3 VNF for this entry-exit pair, creating a path $(1, 2, 4, 3, ...)$. We compare paths created using the two approaches and choose the better one requring less new VNF instances to be created to serve predicted workload.

 
\textbf{Stage Placement Algorithm:} Given the shortest datacenter-to-datacenter path computed in line 19, we decide the optimal placement for deploying VNFs of the service chain onto the datacenters in the path, which minimizes the number of new VNF instances that need to be created. The minimum overall number of new instances to be created is decided using a dynamic programming approach. Let $N(i, j)$ denote the total number of new instances of stage $1$ to stage $j$ VNFs, that should be created to serve the predicted workload $Q_p$ for entry-exit pair $p$, if stage $j$ is placed on the $i$th datacenter in the datacenter-to-datacenter path (suppose the path contains k datacenters and service chain path contains $m$ stages). We have

\begin{equation}
N(1, 1) = num(1,1), N(2, 1) = num(2,1),\nonumber\\
\end{equation}
\begin{equation}
N(i, 1) = +\infty,\forall i= 3,\ldots,k,\nonumber\\
\end{equation}
\begin{equation}
N(1, j) =  N(1, j-1)+num(1,j),\forall j=2,\ldots,m,\nonumber\\
\end{equation}
\begin{equation}
\resizebox{\columnwidth}{!}{$N(i, j) = min\{ N(i-1, j-1)+num(i, j), N(i, j-1)+num(i,j) \}$},\nonumber\\
\end{equation}
\begin{equation}
~~~~~~~~~~~~~~~~~~~~~~~~~~~\forall i=2,\ldots,k, j=2,\ldots,m
\label{eq1}
\end{equation}


\noindent where $num(i, j)$ is the number of new stage-$j$ VNF instances that need to be created when stage $j$ is placed on the $i$th datacenter, {\em i.e.},
\begin{equation}
\resizebox{\columnwidth}{!}{
$num(i,j)=
  \begin{cases}
    0,                                        &\text{if $Q'_{i,j} \geq Q_p$ and $m-j+1 \geq k-i$}\\
    \lceil(Q_p-Q'_{i,j})/C_j\rceil, &\text{if $Q'_{i,j}<Q_p$ and $m-j+1 \geq k-i$}\\
    +\infty,                                &\text{if $m-j+1<k-i$}
  \end{cases}$}
 \label{eq2}
 \end{equation}
 
\noindent Here $Q'_{i,j}$ is the total available capacity of stage-$j$ VNF instances in the $i$th datacenter and $C_j$ is the processing capacity of one stage-$j$ VNF instance. 


The rationale behind Eqn.~(\ref{eq1}) and Eqn.~(\ref{eq2}) is as follows. If stage $j$ is placed on the $i$th datacenter, then stage $j-1$ can only be placed on the $i-1$th datacenter (requiring $N(i-1, j-1)+num(i, j)$ new VNF instances), or on the $i$th datacenter (requiring $N(i, j-1)+num(i,j)$ new VNF instances). Otherwise, the service chain path violates the restrictions given at the beginning of Sec.~\ref{sec:pathcomp}. When $m-j+1<k-i$, {\em i.e.}, the remaining number of stages in the service chain is smaller than the remaining number of datacenters in the shortest path (it is not compulsory to place stage $m$ VNF on exit datacenter $k$, that's why we have a plus 1 on left side of inequation), the stage placement is not  feasible ($num(i,j)$ is set to $+\infty$) since restriction (2) is violated; otherwise, the number of new stage-$j$ VNF instances in datacenter $i$ is computed according to how much additional capacity is needed to serve the predicted workload. 

After the algorithm computes up to $N(k, m)$, we can find the optimal service chain path by tracing back from the smaller of $N(k, m)$ and $N(k-1, m)$, to either $N(1, 1)$ and $N(2, 1)$.

\subsection{Flow Routing On Data Plane}
\label{sec:flow-routing-on-dp}



We next illustrate how DP flow routing is done. 
We use a DP traffic flow sent by the caller as an example. The same routing process can be applied to DP traffic flow sent by the callee.

\noindent\textbf{Enter the Service Chain Path.} The caller learns an IP address located on his entry datacenter when the SIP INVITE transaction ends (see Sec.~\ref{sec:message-routing-on-control-plane}). The caller uses this IP address to send DP traffic flow to his entry datacenter.

\noindent\textbf{Follow the Service Chain Path.} When a flow enters a datacenter, an OpenFlow Packet\_IN message is sent to the local controller. In the entry datacenter, the local controller maps the caller's source IP and source port to the callee's source IP using the first mapping in Table~\ref{mappings}. Using the location service, the local controller also gets to know the exit datacenter of this flow and can then identify the service chain path of the entry-exit datacenter pair for the current scaling interval. In order for local controllers in  subsequent datacenters in the service chain path to learn about the service chain path, the local controller in the entry datacenter adds a tag to the header of packets in the flow (encoded using the destination port field in our implementation). The tag contains the index of the flow's entry datacenter, the index of the flow's exit datacenter and the current scaling interval number modulo 4. Since a local controller only needs to check whether the encoded scaling interval is the previous interval, the current interval or the next interval, we encode the current scaling interval number modulo 4 (only two bits are needed to represent the scaling interval). Local controllers in subsequent datacenters learn the service chain paths for each entry-exit pair as in step 4 of  Sec.~\ref{sec:CP_proactivescale}. 

\noindent\textbf{Choose VNF instances.} After the local controller in a datacenter learns the service chain path that a flow should follow, it selects VNF instances for each stage of the flow's service chain, that is deployed in the datacenter, for the flow to traverse, according to a smallest-workload-first criteria. It enables the flow to go through selected VNF instances by encoding routing information in the flow's packet header too (using the destination IP address field in our implementation).

In particular, a unique index is assigned to each VNF instance in a datacenter. The smallest index value is the number of datacenters that \textit{ScalIMS} controls (the index values from 0 to datacenter number minus 1 is reserved to be used by inter-datacenter VxLAN tunnel). \textit{ScalIMS} configures each VNF instance with 2 NICs, an entry NIC for accepting ingress traffic and an exit NIC for sending egress traffic. Each NIC is connected to an independent virtual switch. A static flow rule is installed on the switch that each VNF instance's entry NIC connects to. In \textit{ScalIMS}, flow rules installed for stage-$i$ $(i=1, ..., 3)$ VNF instances match selected bits in the destination IP address with a mask of $255<<8*(4-i)$. If the matched bits equal the index of a VNF instance, then the flow is sent to the corresponding VNF instance. 

In \textit{ScalIMS}, if the flow has passed through stage $i$ $(i=1, ..., 3)$ VNF and is going to another datacenter for further processing, we fill the matched bits with a mask of $255<<8*[4-(i+1)]$ by the index of the next datacenter. Note that the mask $255<<8*[4-(i+1)]$ is the mask for stage $i+1$ ( If $i=3$, we treat stage $4$ as an virtual exit stage that is always deployed on exit datacenter, with details in first difference at start of Sec.~\ref{sec-dp-scaling}). When the flow comes out of the exit NIC of the last processing stage in the current datacenter, a pre-installed static rule automatically routes the flow to the next datacenter through a VxLAN tunnel~\cite{vxlan}, if the matched bits indicate the index of the next datacenter.

\noindent\textbf{Exit from the DP Service Chain:} When the flow has been processed by VNFs of all stages, 
 it exits from the DP service chain path. To send the flow to the callee, the local controller in the exit datacenter maps the caller's source IP and source port to the callee's entry IP, callee IP, callee receive port according to the fourth mapping in Table~\ref{mappings}.  Then an address translation is carried out to substitute the caller's source IP by the callee's IP, the caller's destination IP by the callee's IP, and the caller's destination port by the callee's receive port. 

\subsection{Handling Scaling Interval Inconsistency}\label{Inconsistency}

In Step 5 of Fig.~\ref{fig:proactive-scaling}, the global controller sends an `enter new scaling interval' message to all local controllers. Given that not all local controllers may receive the message at the same time, there can be temporary inconsistency of the scaling intervals each local controller is currently in. Hence when a local controller 
processes a flow, it may find that the encoded scaling interval in the packet header indicates the previous scaling interval, the current scaling interval, or the next scaling interval.  
For example, suppose local controller 1 has received `enter new scaling interval' message and local controller 2 has not received it. 
Local controller 1 becomes 1 scaling interval ahead of local controller 2. Suppose that a flow enters the service chain from local controller 1's datacenter and is immediately sent to local controller 2's datacenter during this short time window. Then local controller 2 will find that the encoded scaling interval in the flow's packes is the next scaling interval. But if flow 1 goes from a reverse direction during this time interval (enter service chain from local controller2's datacenter and be sent to to local controller 1's datacenter), then local controller 1 will find that flow 1's encoded scaling interval is the previous scaling interval. 

To counter this problem, 
three sets of service chain paths are always retained in each local controller during the execution of proactive scaling, for the previous, current and next scaling intervals, respectively. 
 When a local controller receives a new flow, it uses one of the service chain paths depending on the scaling interval encoded in the header of the flow packets.
 

%% file: evaluation.tex
\section{Implementation and Evaluation} \label{Evaluation}

\subsection{Implementation}

We implement a prototype of \textit{ScalIMS} in Java. The local controller is implemented as a module in the FloodLight SDN controller~\cite{floodlight}. The global controller is implemented as a multi-threaded java server program, which communicates with local controllers over regular sockets.  We also implement a traffic generator based on PJSIP~\cite{pjsip}. We deploy one traffic generator in each datacenter, producing user arrivals bond to the datacenter at a configurable rate. A global traffic generator coordinator receives notifications of generated users and pairs up users in a first-come-first-match manner. A call process is launched between every pair of paired users, which includes all the required SIP transactions to establish a call, followed by a one-minute voice call at the bit rate of 80kbit/s, as well as necessary SIP transactions to shutdown the call.



We use P-CSCF, S-CSCF and HSS components from the Clearwater Project~\cite{project-clearwater} as our control plane VNFs. The data plane service chain contains a firewall (implemented using user space Click~\cite{martins2014clickos}), an intrusion detector (Snort IDS ~\cite{snort}) and a transcoder (implemented using user space Click). Each network function runs on a QEMU/KVM virtual machine (VM). A CP VM is configured with 1 core and 2GB RAM and a DP VM is configured with 2 cores and 2GB RAM. The capacity and overload threshold (to decide scaling out) of each instance of each VNF are shown in Table~\ref{table:stat}, which are obtained by stress testing VNF instances to overloaded states. 

\begin{table}[!t]
\centering
\caption{VNF Capacity, Overload Threshold and Inter-datacenter Delays}
\label{table:stat}
\resizebox{\columnwidth}{!}{
\begin{tabular}{|l|l|l|l|l|llllll}
\cline{1-5} \cline{7-11}
VNF        & Capacity    & \begin{tabular}[c]{@{}l@{}}CPU\\ threshold\end{tabular} & \begin{tabular}[c]{@{}l@{}}Memory\\ threshold\end{tabular} & \begin{tabular}[c]{@{}l@{}}Input pkt/s\\ threshold\end{tabular} & \multicolumn{1}{l|}{} & \multicolumn{1}{l|}{\begin{tabular}[c]{@{}l@{}}Inter-Group\\ Delay\end{tabular}} & \multicolumn{1}{l|}{$g_0$} & \multicolumn{1}{l|}{$g_1$} & \multicolumn{1}{l|}{$g_2$} & \multicolumn{1}{l|}{$g_3$} \\ \cline{1-5} \cline{7-11}
P-CSCF     & 500 tran/s  & 70\%                                                    & 50\%                                                       & 1000 pkt/s                                                      & \multicolumn{1}{l|}{} & \multicolumn{1}{l|}{$g_0$}                                                 & \multicolumn{1}{l|}{0ms}   & \multicolumn{1}{l|}{10ms}  & \multicolumn{1}{l|}{15ms}  & \multicolumn{1}{l|}{50ms}  \\ \cline{1-5} \cline{7-11}
S-CSCF     & 200 tran/s  & 70\%                                                    & 50\%                                                       & 400 pkt/s                                                       & \multicolumn{1}{l|}{} & \multicolumn{1}{l|}{$g_1$}                                                 & \multicolumn{1}{l|}{10ms}  & \multicolumn{1}{l|}{0ms}   & \multicolumn{1}{l|}{20ms}  & \multicolumn{1}{l|}{17ms}  \\ \cline{1-5} \cline{7-11}
Firewall   & 35000 pkt/s & 90\%                                                    & 50\%                                                       & 35000 pkt/s                                                     & \multicolumn{1}{l|}{} & \multicolumn{1}{l|}{$g_2$}                                                 & \multicolumn{1}{l|}{15ms}  & \multicolumn{1}{l|}{20ms}  & \multicolumn{1}{l|}{0ms}   & \multicolumn{1}{l|}{15ms}  \\ \cline{1-5} \cline{7-11}
IDS        & 20000 pkt/s & 90\%                                                    & 50\%                                                       & 2000 pkt/s                                                      & \multicolumn{1}{l|}{} & \multicolumn{1}{l|}{$g_3$}                                                 & \multicolumn{1}{l|}{50ms}  & \multicolumn{1}{l|}{17ms}  & \multicolumn{1}{l|}{15ms}  & \multicolumn{1}{l|}{0ms}   \\ \cline{1-5} \cline{7-11}
Transcoder & 15000 pkt/s & 90\%                                                    & 50\%                                                       & 15000 pkt/s                                                     &                       &                                                                            &                            &                            &                            &                            \\ \cline{1-5}
\end{tabular}
}
\end{table}

\subsection{Evaluation in IBM SoftLayer Cloud}

We evaluate the performance of \textit{ScalIMS} on IBM SoftLayer Cloud~\cite{softlayer} by deploying 4 bare-metal servers in 4 Softlayer datacenters, located in Tokyo, Hong Kong, London and Houston, respectively. Each server is equipped with two 6-core 2.4GHz Intel CPU, 64GB RAM, 1TB SATA disk. In the SoftLayer cloud, servers in different datacenters are connected through a global private network~\cite{sl-network}, which provides a Gigabyte throughput. \textit{ScalIMS} creates a VxLAN tunnel mesh in the private network to route DP traffic. CP VNFs are connected directly over the private network.   


 We use two groups of metrics to evaluate the performance of \textit{ScalIMS}. (1) The total number of new VNF instances created over time: the smaller the number is, the more cost/resource effective \textit{ScalIMS} is; (2) User traffic quality metrics: the SIP transaction completion time for the control plane traffic, and the RTT and loss rate for data plane traffic. 

Each traffic generator is configured to start generating users at a small rate, which gradually increases to a maximal rate, and then gradually decreases. The user generation rate is changed 
in every \textbf{change interval}. Each scaling interval may contain multiple change intervals.
The scaling interval is set to be 50 seconds long and each buffer queue retains an unused VNF instance for at most 10 scaling intervals. The maximum allowed end-to-end flow delay is set to 250ms.

In the following experiments, we compare the performance achieved by using proactive scaling only (the local controller does not react to overload of instances), reactive scaling only (the global controller initializes a good set of service chain paths, but no subsequent proactive scaling decisions are broadcast to local controllers), and both proactive and reactive scaling enabled ({\em i.e.}, the combined scaling stragegy of \textit{ScalIMS}).

\subsubsection{Simultaneous Start}

In this set of experiments, each traffic generator starts generating users simultaneously at the rate of $1~users/s$, which gradually increases to $15 users/s$ and then decreases to $6 users/s$.  
In this way, the peak workload arrives almost concurrently in each datacenter. The duration of a change interval is set to 20s, 30s, ... 80s respectively.

\begin{figure}[!t]
        \centering
        \includegraphics[width=\columnwidth]{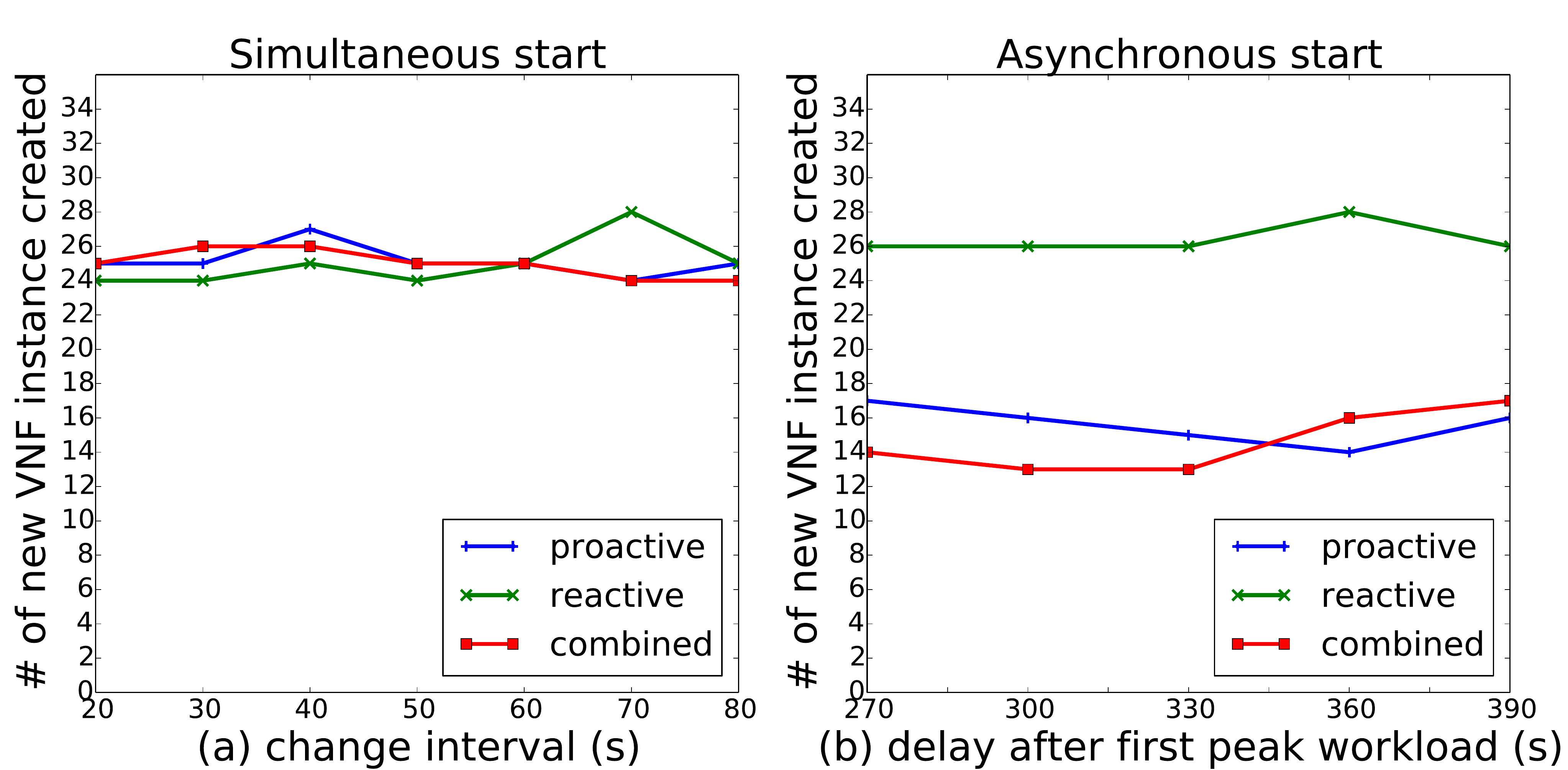}
        \caption{Overall number of new VNF instances created. (a) Simultaneous start. (b) Asynchronous start.}
        \label{fig:nf-creation}
\end{figure}

\begin{figure}[!t]
 \centering
        \includegraphics[width=\columnwidth]{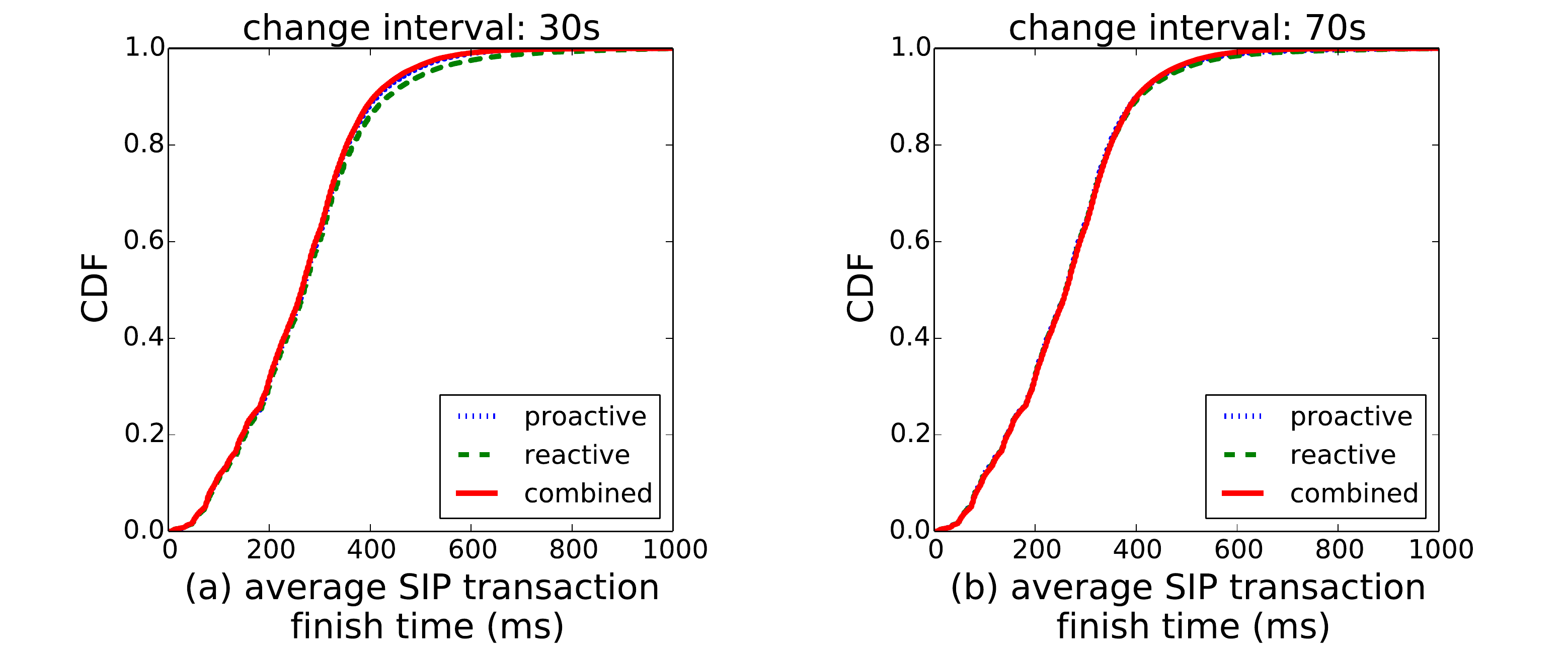}
        \caption{CP traffic quality with simultaneous start.}
        \label{fig:syn-cp}
\end{figure}

\begin{figure}[!t]
        \centering
        \includegraphics[width=\columnwidth]{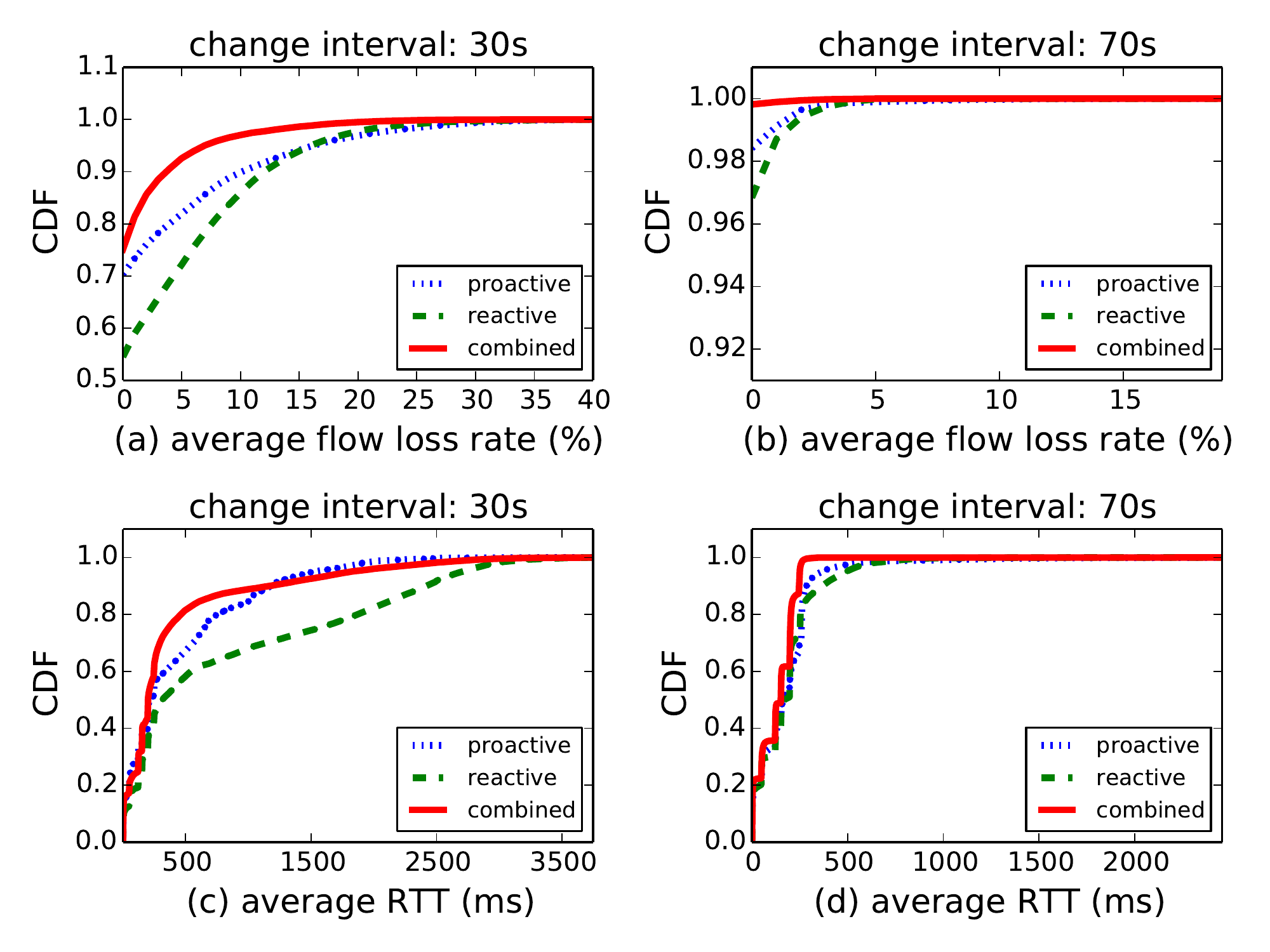}
        \caption{DP traffic quality with simultaneous start.}
        \label{fig:syn-lossrate-rtt}
\end{figure}

Fig.~\ref{fig:nf-creation}(a) and Fig.~\ref{fig:syn-cp} show that the total number of VNF instances provisioned and the average SIP transaction completion time under the three schemes do not differ much. The total number of VNF instances provisioned is similar because the maximum workload on each datacener is similar. 
On the other hand, Fig.~\ref{fig:syn-lossrate-rtt} shows that the combined scaling strategy of {\em ScalIMS} out-performs the other two strategies in terms of data plane traffic quality. 
The reason can be explained as follows. Pure reactive scaling adds new instances only when an overload signal arises. During the boot-up time of new instances, traffic continues to arrive at the overloaded VNF instances, resulting in a high packet loss rate and then high RTT. Pure proactive scaling adjusts VNF instances once every scaling interval. During each scaling interval, increased workload may have overloaded the system. The best performance is achieved combining both proactive and reactive scaling. The reason for similar CP performance is that scaling in the CP service chains is not triggered as often as that of the DP service chains, since each instance of a CP VNF is able to handle a large number of SIP transactions. 




\subsubsection{Asynchronous Start}

In this set of experiments, traffic generators are not started at the same time. Traffic generator in the Tokyo datacenter is started first, followed by traffic generator in the Hong Kong datacenter, in the Houston datacenter and then in the London datacenter. The start time of the traffic generator in a subsequent datacenter is delayed for a configurable period of time after that of the previously started traffic generator. In this way, the peak workloads in different datacenter do not occur at the same time. Each traffic generator increases its user generation rate from 5 $users/s$ to 15 $users/s$ and then dereases it to $5 users/s$, with a 30s change interval.

\begin{figure}[!t]
 \centering
        \includegraphics[width=\columnwidth]{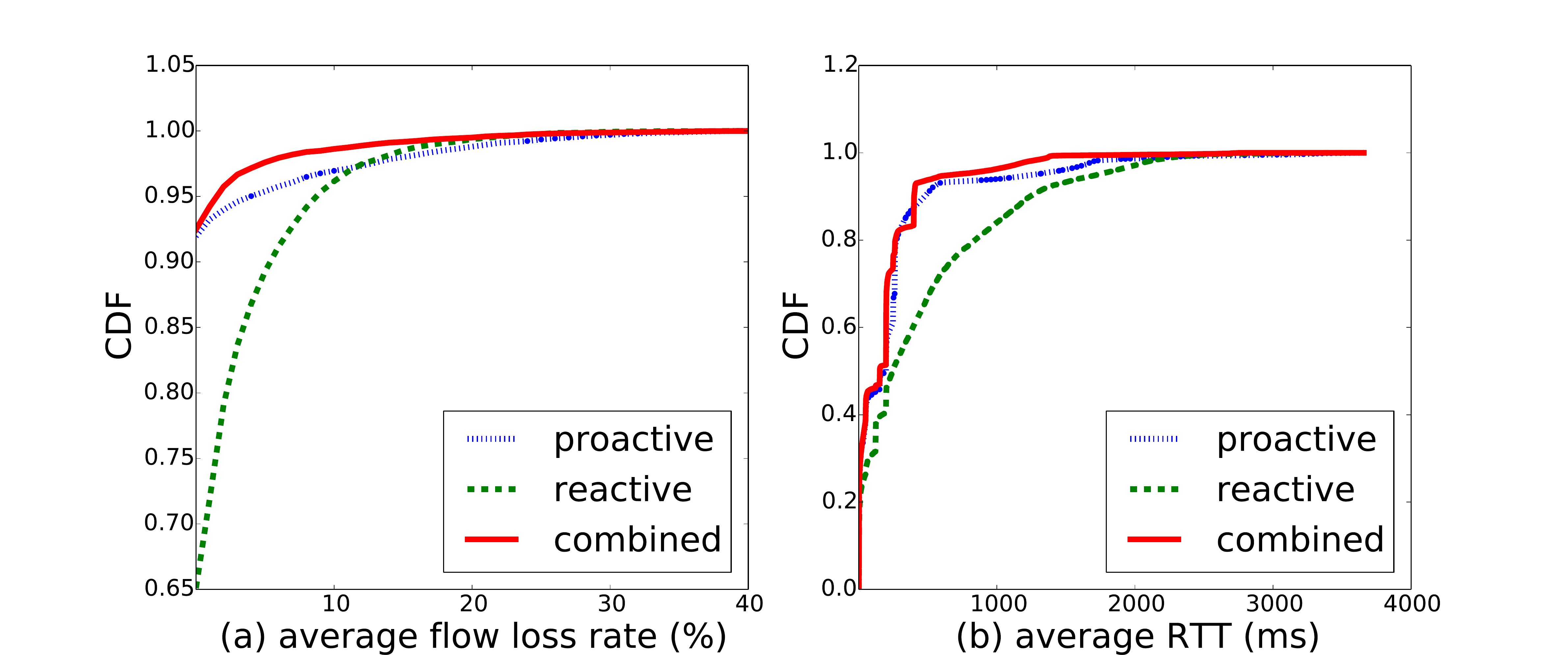}
        \caption{DP traffic quality with asynchronous start.}
        \label{fig:nonsyn-rttloss}
\end{figure}

In Fig.~\ref{fig:nf-creation}(b), the start delay between traffic generators in Tokyo $\sim$ and in Hong Kong is set according to values in the x axis, while start delays between Hong Kong and Houston and between Houston and London are set to 120s. The figure shows that the number of VNF instances created by the combined scaling approach is always the smallest. Fig.~\ref{fig:nonsyn-rttloss} shows that the traffic quality is much better with the combined approach as well. 

\begin{figure}[!t]
        \centering
        \includegraphics[width=1\columnwidth]{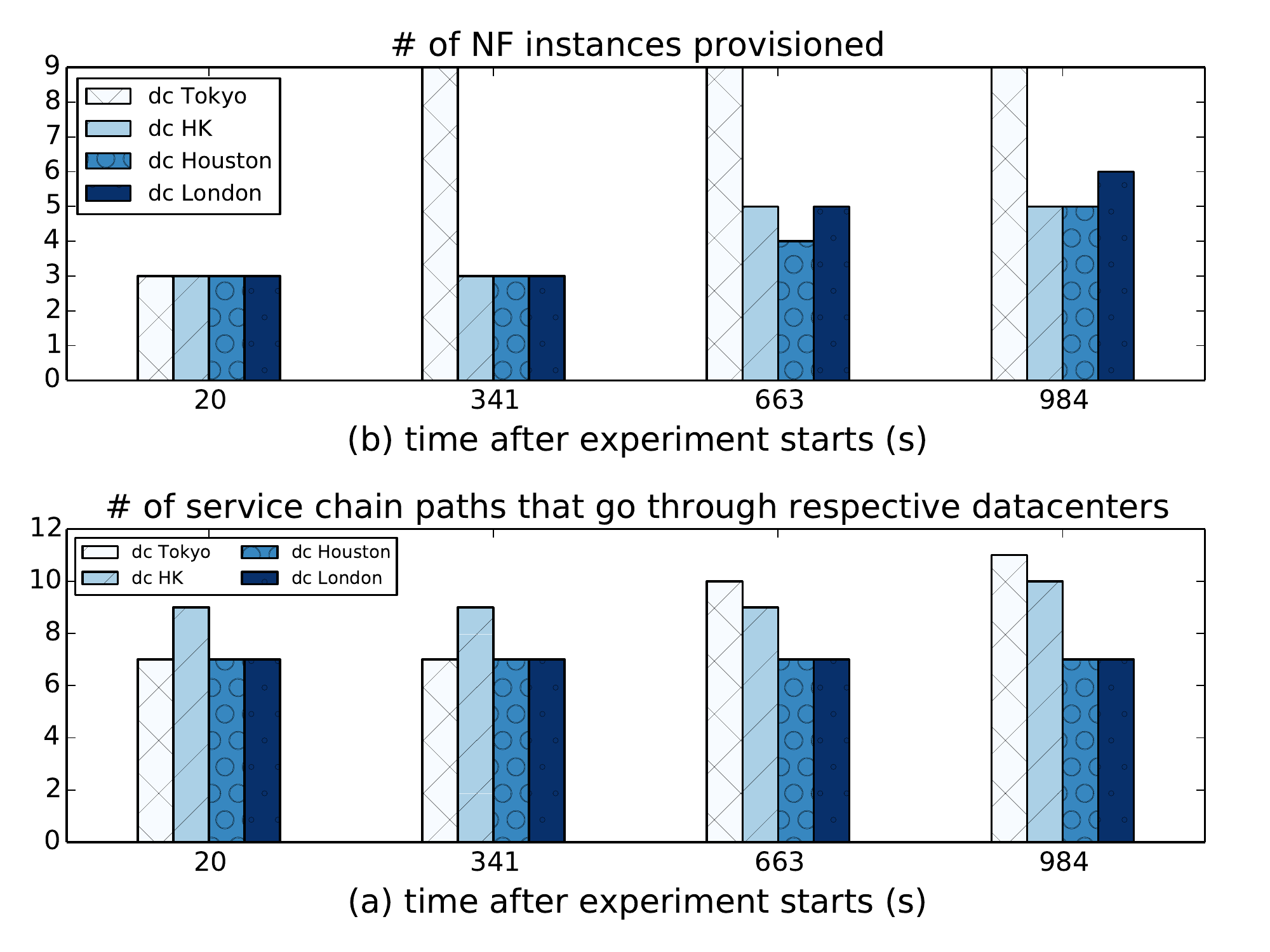}
        \caption{
		(a) Number of provisioned VNF instances in each datacenter verses time. (b) The number of service chain paths that go through different datacenters.}
        \label{fig:nonsyn-scpath}
\end{figure}

Why can the combined scaling perform well even when it creates a smaller number of VNF instances? The Tokyo datacenter sees the peak workload first, leading to the provisioning of many VNF instances in the datacenter, which later become redundant and will be buffered for 10 scaling intervals. When peak workload arrives at other datacenters, the global controller can re-use the existing VNF instances by creating service chain paths traversing Tokyo datacenter, increasing resource utilization efficiency. 
Fig.~\ref{fig:nonsyn-scpath}(a) shows the number of VNF instances created in each datacenter verses the time after the experiment starts. We can see that due to the early arrival of peak workload, many VNF instances are provisioned in Tokyo datacenter by 341s. After that time, the number of service chain paths that go through Tokyo datacenter increases from 7 to 11 (Fig.~\ref{fig:nonsyn-scpath}(b)) because the global controller schedules more DP traffic flows to re-use these existing VNF instances. Because these existing VNF instances are buffered and reused, DP traffic flows do not go through overloaded instances, improving the quality experienced by the DP flows.

\subsection{Evaluation in Emulation Cluster}

\subsubsection{Links with Large Delays}

Next we evaluate whether {\em ScalIMS} is able to effectively detect links with high delays and shift the traffic away from these links on our own emulation cluster. We use 8 servers in the cluster and emulate inter-datacenter delay with Linux TC. The 8 datacenters are separated into 4 groups (i.e. $g_0$ to $g_3$, where $g_i$ contains datacenter $i$ and $i+1$). The delay between datacenters in the same group is set to 5ms, the delay between datacenters in different group is set according to Table~\ref{table:stat}.

In this set of experiments, traffic generators increase user production rate from 1 $users/s$ to 12 $users/s$ and then decrease the rate to 6 $users/s$, with a 50s update interval. At 350s after the experiments start, we deliberately increase the delay between groups $1$ and $3$ to 65ms. Since the maximum end-to-end delay threshold is 50ms in our DP proactive scaling algorithm, the global controller would avoid using links between groups $1$ and $3$. 

\begin{figure}[ht]
        \centering
        \includegraphics[width=1\columnwidth]{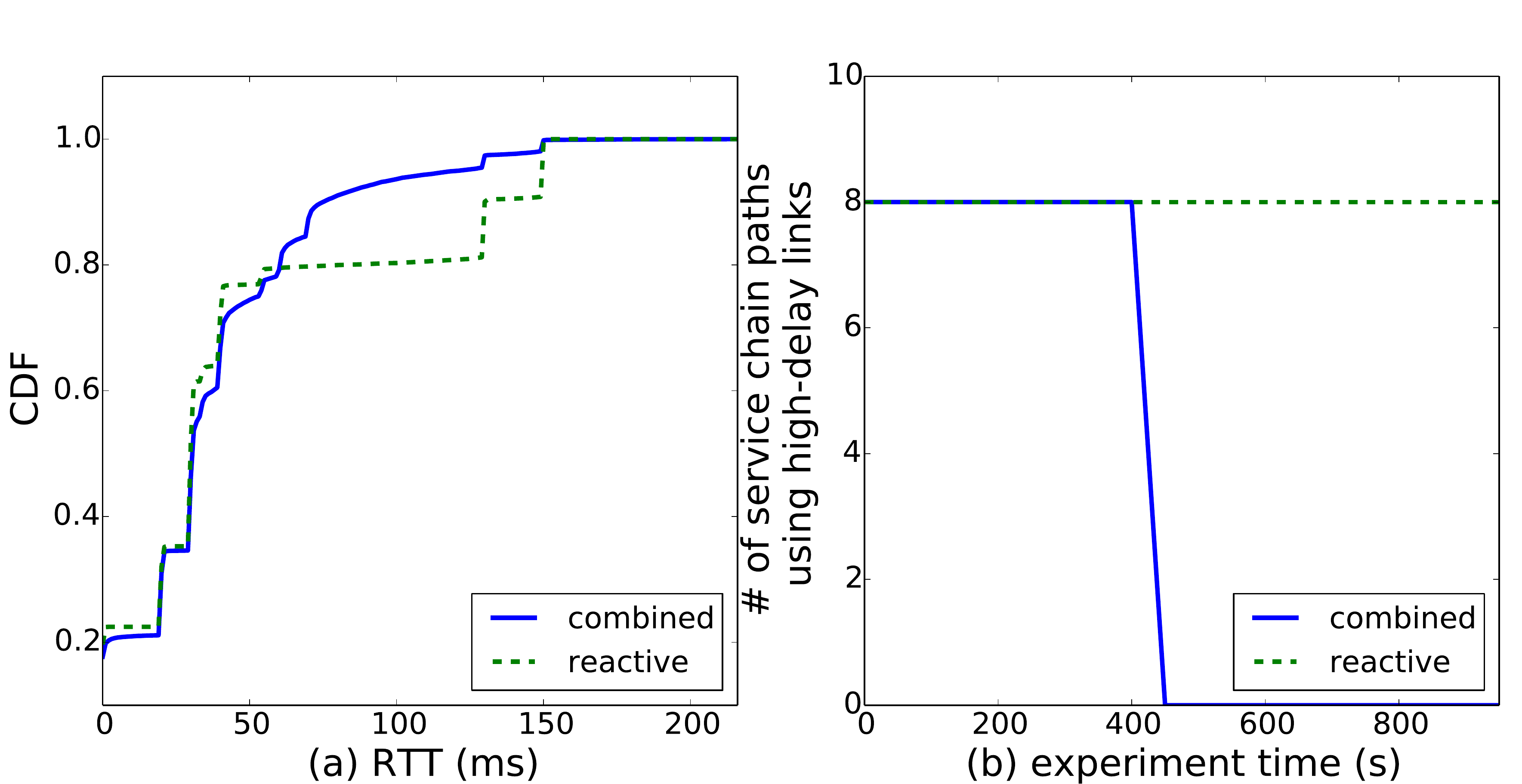}
        \caption{(a) CDF of flow RTT. (b) Number of service chain paths that use the links between group 1 and group 3.}
        \label{fig:delayrun}
\end{figure}

Fig.~\ref{fig:delayrun} (a) shows that with {\em ScalIMS}, the number of flows whose measured RTT is smaller than 100ms is 15\% more than that with reactive scaling. 
Fig.~\ref{fig:delayrun}(b) illustrates that after the global controller detects the links with high delay, it re-computes new service chain paths that avoid using those links. At about 400s, the number of service chain paths that use the high delay links drops from 8 to 0, boosting the quality of the DP media flow.

\subsubsection{Scaling Interval Consistency}

We also evaluate the impact of our design on maintaining scaling interval consistency as discussed in Sec.~\ref{Inconsistency}, by comparing it with one where flows are not tagged with scaling intervals and local controllers always route flows using servcice chain paths in the current scaling interval. In this set of experiments, we generate 15 calls per second between datacenter 0 and 1 for 5 minutes. Enough VNF instances are pre-provisioned on 3 datacenters (datacenters $0-2$) to serve the traffic. The global controller alternates service chain paths from datacenter 0 to datacenter 1 and from datacenter 1 to datacenter 0 from (0, 0, 0, 1, 1) and (1, 1, 1, 0, 0) to (0, 2, 2, 2, 1) and (1, 2, 2, 2, 0) in each scaling interval. 
The delays between datacenters 0 and 2 and between datacenters 1 and 2 are both set to 400ms.
If a local controller finds out that it is not on a flow's service chain path, it applies a rule that drops all the packets of this flow. 


We find that without scaling interval tagging, 18 out of 8722 flows ($0.2\%$) are completely not received by their receivers. On the contrary, all 8724 flows are received with 0\% loss rate when the scaling interval tagging is enabled. This is because when service chain paths from datacenter 0 to datacenter 1 and from datacenter 1 to datacenter 0 are switched from (0, 2, 2, 2, 1) and (1, 2, 2, 2, 0) to (0, 0, 0, 1, 1) and (1, 1, 1, 0, 0), datacenter 2 enters the next scaling interval earlier. There is a small time window when datacenter 2 starts using service chain path (0, 0, 0, 1, 1) and (1, 1, 1, 0, 0) while datacenter 0 and datacenter 1 are still in the previous scaling interval, using service chain path (0, 2, 2, 2, 1) and (1, 2, 2, 2, 0). Datacenters 0 and 1 continue sending flows to datacenter 2 during that time window, but datacenter 2 finds that it is not on the service chain path of this flow, thus dropping all the packets. We omit the related figures due to space constraint.

%% file: conclusion.tex
\section{Conclusion}

Scaling VNF service chains has become an appealing topic in this era. In this paper, we propose \textit{ScalIMS}, a scaling system designed to operate in geo-distributed datacenter for scaling IMS system. We evaluate our prototype implementation on both IBM SoftLayer cloud and am emulated cluster. The evaluation result shows that: \textit{First}, \textit{ScalIMS} is able to improve user traffic quality by a large margin when compared with pure reactive scaling approach. \textit{Second}, when peak workload arrives asynchronously, \textit{ScalIMS} is able to decrease total number of VNF instances while providing excellent user traffic quality. \textit{Third}, \textit{ScalIMS} is very efficient in avoiding inter-datacenter network link with large delays. \textit{Finally}, we show that distributed routing framework of \textit{ScalIMS} accurately route flows across geo-distributed datacenters.